\documentclass[fleqn,10pt,twocolumn]{wlscirep}

\usepackage[utf8]{inputenc}
\usepackage[T1]{fontenc}
\usepackage[switch]{lineno} 

\usepackage{upgreek}

\usepackage{framed}

\usepackage{changepage}
\usepackage{colortbl}
\usepackage{marvosym}

\usepackage{sansmath}
\makeatletter
\providecommand*\AtBeginEnvironment[1]{%
  \@ifundefined{#1}%
    {\@latex@error{Environment #1 undefined}\@ehc
     \@gobble}%
    {\@ifundefined{ABE@env@#1}%
       {\expandafter\let\csname ABE@env@#1\expandafter\endcsname
          \csname #1\endcsname
        \expandafter\let\csname ABE@hook@#1\endcsname\@empty
        \@namedef{#1}{\@nameuse{ABE@hook@#1}\@nameuse{ABE@env@#1}}}%
       {}%
     \expandafter\g@addto@macro\csname ABE@hook@#1\endcsname}}
\@onlypreamble\AtBeginEnvironment
\makeatother

\DeclareSymbolFont{Greekletters}{OT1}{iwona}{m}{n}
\DeclareSymbolFont{greekletters}{OML}{iwona}{m}{it}
\DeclareMathSymbol{\Phi}{\mathord}{Greekletters}{"08}

\let\oldtabular\tabular
\renewcommand{\tabular}{\small\sffamily\oldtabular}

\usepackage{tikz}

\usepackage{pgfplots}
\pgfplotsset{compat=newest}

\makeatletter
\pgfmathdeclarefunction{erf}{1}{%
  \begingroup
    \pgfmathparse{#1 > 0 ? 1 : -1}%
    \edef\sign{\pgfmathresult}%
    \pgfmathparse{abs(#1)}%
    \edef\x{\pgfmathresult}%
    \pgfmathparse{1/(1+0.3275911*\x)}%
    \edef\t{\pgfmathresult}%
    \pgfmathparse{%
      1 - (((((1.061405429*\t -1.453152027)*\t) + 1.421413741)*\t 
      -0.284496736)*\t + 0.254829592)*\t*exp(-(\x*\x))}%
    \edef\y{\pgfmathresult}%
    \pgfmathparse{(\sign)*\y}%
    \pgfmath@smuggleone\pgfmathresult%
  \endgroup
}

\usetikzlibrary{arrows}
\usepgfplotslibrary{fillbetween}
\usetikzlibrary{calc}
\usetikzlibrary{math}
\usetikzlibrary{calc}
\usetikzlibrary{patterns}
\usetikzlibrary{patterns.meta}
\usetikzlibrary{arrows.meta}
\usetikzlibrary{arrows}
\usetikzlibrary{positioning}
\usetikzlibrary{shapes, shapes.geometric, arrows}
\usetikzlibrary{arrows.meta}
\usetikzlibrary{patterns,patterns.meta}
\usetikzlibrary{intersections}
\usetikzlibrary{backgrounds}
\usepackage{pgfplotstable}
\usepgfplotslibrary{patchplots}
\pgfplotsset{
tick label style = {font=\sansmath\sffamily},
every axis label/.append style = {font=\sansmath\sffamily}
}
\def\centerarc[#1](#2)(#3:#4:#5)
{ \draw[#1] ($(#2)+({#5*cos(#3)},{#5*sin(#3)})$) arc (#3:#4:#5); }

\pgfplotsset{
    every first x axis line/.style={},
    every first y axis line/.style={},
    every first z axis line/.style={},
    every second x axis line/.style={},
    every second y axis line/.style={},
    every second z axis line/.style={},
    first x axis line style/.style={/pgfplots/every first x axis line/.append style={#1}},
    first y axis line style/.style={/pgfplots/every first y axis line/.append style={#1}},
    first z axis line style/.style={/pgfplots/every first z axis line/.append style={#1}},
    second x axis line style/.style={/pgfplots/every second x axis line/.append style={#1}},
    second y axis line style/.style={/pgfplots/every second y axis line/.append style={#1}},
    second z axis line style/.style={/pgfplots/every second z axis line/.append style={#1}}
}

\pgfdeclareplotmark{rottriangle}{%
  \pgfpathmoveto{\pgfqpoint{0pt}{-\pgfplotmarksize}}%
  \pgfpathlineto{\pgfqpointpolar{30}{\pgfplotmarksize}}%
  \pgfpathlineto{\pgfqpointpolar{150}{\pgfplotmarksize}}%
  \pgfpathclose
  \pgfusepathqstroke
}%
\pgfdeclareplotmark{rottriangle*}{%
  \pgfpathmoveto{\pgfqpoint{0pt}{-\pgfplotmarksize}}%
  \pgfpathlineto{\pgfqpointpolar{30}{\pgfplotmarksize}}%
  \pgfpathlineto{\pgfqpointpolar{150}{\pgfplotmarksize}}%
  \pgfpathclose
  \pgfusepathqfillstroke
}

\pgfdeclareplotmark{halffdiamondtop*}{%
  \pgfpathmoveto{\pgfqpoint{0pt}{\pgfplotmarksize}}
  \pgfpathlineto{\pgfqpoint{0.7\pgfplotmarksize}{0pt}}
  \pgfpathlineto{\pgfqpoint{0pt}{-\pgfplotmarksize}}
  \pgfpathlineto{\pgfqpoint{-0.7\pgfplotmarksize}{0pt}}
  \pgfpathclose
  \pgfusepathqstroke
  \pgfpathmoveto{\pgfqpoint{0pt}{\pgfplotmarksize}}
  \pgfpathlineto{\pgfqpoint{0.7\pgfplotmarksize}{0pt}}
  \pgfpathlineto{\pgfqpoint{-0.7\pgfplotmarksize}{0pt}}
  \pgfpathclose
  \pgfusepathqfill
  \pgfusepathqstroke
}

\pgfdeclareplotmark{halffilledsquare}{%
  \pgfpathrectangle{\pgfpoint{-\pgfplotmarksize}{-\pgfplotmarksize}}%
                   {\pgfpoint{2\pgfplotmarksize}{2\pgfplotmarksize}}%
  \pgfusepathqstroke
  \pgfpathrectangle{\pgfpoint{-\pgfplotmarksize}{0pt}}%
                   {\pgfpoint{2\pgfplotmarksize}{\pgfplotmarksize}}%
}

\pgfdeclareplotmark{halffilledsquare*}{%
  \pgfpathrectangle{\pgfpoint{-\pgfplotmarksize}{0pt}}%
                   {\pgfpoint{2\pgfplotmarksize}{\pgfplotmarksize}}%
  \pgfusepathqfill
  \pgfpathrectangle{\pgfpoint{-\pgfplotmarksize}{-\pgfplotmarksize}}%
                   {\pgfpoint{2\pgfplotmarksize}{2\pgfplotmarksize}}%
  \pgfusepathqstroke
}

\tikzset{
  font={\small\sffamily\sansmath}
}

\pgfplotsset{every tick label/.append style={font=\small\sffamily\sansmath}}
\pgfplotsset{every tick label/.append style={font=\small}}

\makeatletter
\def\pgfplots@drawaxis@outerlines@separate@onorientedsurf#1#2{%
    \if2\csname pgfplots@#1axislinesnum\endcsname
    \else
    \scope[/pgfplots/every outer #1 axis line,
        #1discont,decoration={pre length=\csname #1disstart\endcsname, post length=\csname #1disend\endcsname}]
        \pgfplots@ifaxisline@B@onorientedsurf@should@be@drawn{0}{%
            \draw [/pgfplots/every first #1 axis line] decorate {
                \pgfextra
                \pgfplotspointonorientedsurfaceabsetupfor{#2}{#1}{\pgfplotspointonorientedsurfaceN}%
                \pgfplots@drawgridlines@onorientedsurf@fromto{\csname pgfplots@#2min\endcsname}%
                \endpgfextra 
                };
        }{}%
        \pgfplots@ifaxisline@B@onorientedsurf@should@be@drawn{1}{%
            \draw [/pgfplots/every second #1 axis line] decorate {
                \pgfextra
                \pgfplotspointonorientedsurfaceabsetupfor{#2}{#1}{\pgfplotspointonorientedsurfaceN}%
                \pgfplots@drawgridlines@onorientedsurf@fromto{\csname pgfplots@#2max\endcsname}%
                \endpgfextra 
                };
        }{}%
    \endscope
    \fi
}%
\makeatother

\DeclareMathVersion{sans}
\SetSymbolFont{operators}{sans}{OT1}{cmbr}{m}{n}
\SetSymbolFont{letters}{sans}{OML}{cmbrm}{m}{it}
\SetSymbolFont{symbols}{sans}{OMS}{cmbrs}{m}{n}
\SetMathAlphabet{\mathit}{sans}{OT1}{cmbr}{m}{sl}
\SetMathAlphabet{\mathbf}{sans}{OT1}{cmbr}{bx}{n}
\SetMathAlphabet{\mathtt}{sans}{OT1}{cmtl}{m}{n}
\SetSymbolFont{largesymbols}{sans}{OMX}{iwona}{m}{n}

\usepackage{csvsimple}
\usepackage{changepage}
\usepackage[tbtags]{mathtools}
\usepackage{siunitx}[=v2]
\DeclareSIUnit\fps{fps}

\usepackage{csquotes}
\usepackage{enumerate}
\usepackage{enumitem}
\usepackage{filecontents}

\usepackage{nicefrac}

\usepackage{cleveref}
\usepackage{comment}

\usepackage{blindtext}
\usepackage{amssymb}

\usepackage{multirow, makecell}
\usepackage{booktabs}
\usepackage{array}
\usepackage{longtable, tabularx}
\newcolumntype{L}[1]{>{\raggedright\let\newline\\\arraybackslash\hspace{0pt}}m{#1}}
\newcolumntype{C}[1]{>{\centering\let\newline\\\arraybackslash\hspace{0pt}}m{#1}}
\newcolumntype{R}[1]{>{\raggedleft\let\newline\\\arraybackslash\hspace{0pt}}m{#1}}
\newcolumntype{Y}{>{\centering\arraybackslash}X}

\usepackage[normalem]{ulem}
\definecolor{brickred}{rgb}{0.8, 0.25, 0.33}

\newcommand\red[1]{%
  {\textcolor{black}{#1}}%
}

\definecolor{brickred}{rgb}{0.8, 0.25, 0.33}%
\definecolor{darkorange}{rgb}{1.0, 0.55, 0.0}%
\definecolor{persiangreen}{rgb}{0.0, 0.65, 0.58}%
\definecolor{persianindigo}{rgb}{0.2, 0.07, 0.48}%
\definecolor{cadet}{rgb}{0.33, 0.41, 0.47}%
\definecolor{turquoisegreen}{rgb}{0.63, 0.84, 0.71}%
\definecolor{sandybrown}{rgb}{0.96, 0.64, 0.38}%
\definecolor{blueblue}{rgb}{0.0, 0.2, 0.6}%
\definecolor{ballblue}{rgb}{0.13, 0.67, 0.8}%
\definecolor{greengreen}{rgb}{0.0, 0.5, 0.0}%
\definecolor{gold}{rgb}{0.86, 0.71, 0.23}%
\definecolor{mediumtealblue}{rgb}{0.0, 0.33, 0.71}%
\definecolor{persianblue}{rgb}{0.11, 0.22, 0.73}
\definecolor{sacramentostategreen}{rgb}{0.0, 0.34, 0.25}
\definecolor{tyrianpurple}{rgb}{0.4, 0.01, 0.24}
\definecolor{trueblue}{rgb}{0.0, 0.45, 0.81}
\definecolor{darkraspberry}{rgb}{0.53, 0.15, 0.34}
\definecolor{carminepink}{rgb}{0.92, 0.3, 0.26}
\definecolor{celadon}{rgb}{0.67, 0.88, 0.69}
\definecolor{charcoal}{rgb}{0.21, 0.27, 0.31}
\definecolor{cadmiumgreen}{rgb}{0.09, 0.45, 0.27}
\definecolor{LightBlueGreen}{RGB}{102, 194, 160}
\definecolor{Orange}{RGB}{253, 174, 97}
\definecolor{caribbeangreen}{rgb}{0.0, 0.8, 0.6}

\usepackage[
	nonumberlist, 
	acronym, 
	nomain,
	nopostdot,
]{glossaries}

\DeclareCiteCommand{\citenum}
  {}
  {\printfield{labelnumber}}
  {}
  {}

\newacronym{isru}{ISRU}{\textit{in situ} resource utilisation}
\newacronym{gtb}{GraviTower}{GraviTower Bremen Pro} 
\newacronym{zarm}{ZARM}{Center of Applied Space Technology and Microgravity}
\newacronym{1g}{$1g$}{Earth gravity}
\newacronym{mug}{$\mu g$}{microgravity}
\newacronym{dem}{DEM}{discrete element methods}
\newacronym{iss}{ISS}{international space station}
\newacronym{afm}{AFM}{atomic force microscopy}
\newacronym{vdw}{vdW}{van der Waals}

\title{Granular clogging across gravities: a unified scaling}

\author[1,*]{Oliver Gaida}
\author[2,3,\Cross]{Olfa D'Angelo}
\author[4,5,1]{Jonathan E.\ Kollmer}

\affil[1]{Universit\"{a}t Duisburg-Essen, Physics Department, Duisburg, Germany}
\affil[2]{Van der Waals-Zeeman Institute, University of Amsterdam, Amsterdam, The Netherlands}
\affil[3]{Institut Sup\'{e}rieur de l'A\'{e}ronautique et de l'Espace (ISAE-SUPAERO), Universit\'{e} de Toulouse, Toulouse, France} 
\affil[4]{Institute for Frontier Materials on Earth and in Space, German Aerospace Center (DLR), Linder H\"{o}he, 51170 Cologne, Germany}
\affil[5]{Institut f\"{u}r Theoretische Physik, University of Cologne, Z\"{u}lpicher Stra\ss{}e 77, 50937 Cologne, Germany}

\affil[*]{oliver.gries@uni-due.de}
\affil[\Cross]{o.dangelo@uva.nl}

\begin{abstract}
Lacking a universal law for granular flows across gravitational environments,
fundamental processes such as hopper discharge remain vulnerable to failure in low gravity environments. 
A central challenge is clogging, the spontaneous arrest of flow through a constriction; 
yet previous studies report contradictory results on its dependence on gravitational acceleration.
We identify the granular Bond number as the missing control parameter, defined as the ratio of intrinsic cohesive interactions among particles to gravity. Based on an in-bulk measurement of this quantity, we propose to rescale Earth-measured data for predicting granular behavior in low gravity.
We present experiments of granular flow through an orifice under true reduced gravity (Moon and Mars), using an active drop tower, and extraterrestrial soil simulants as model cohesive materials. 
Our experiments reveal substantially increases in clogging probability, contrary to previously predicted,
which depends on the properties of the material itself.
When rescaled by the Bond number, seemingly conflicting results can be explained and collapse into a unified state diagram, predicting clogging across materials and gravitational accelerations. 
This establishes a general framework for the cohesion-to-gravity competition. 
Future space missions to the Moon, Mars, and asteroids will rely on such predictions of granular behavior in low gravity.

\vspace{6pt}
\end{abstract}

\begin{document}
\flushbottom
\maketitle

\begin{framed}\sffamily\small
\noindent
\textbf{On the Moon, will a hopper filled with sand flow or clog?} 
The seemingly simple question of clogging has challenged engineers for centuries -- and becomes even more complex when gravity itself changes. As humanity prepares for missions to the Moon and beyond, understanding how (low) gravity influences granular matter becomes critical. We present a predictive framework to scale granular flow from Earth to any gravitational environment. Through true low gravity experiments, we show that it dramatically increases clogging probability. By quantifying this shift, we introduce a scaling law that unifies our results into a single state diagram. Our findings 
explain previously contradicting results and propose
a foundation for predicting and controlling granular flow in space.
\end{framed}

\section*{Introduction}

Granular materials are everywhere in our Solar System, 
from planetary regolith to rubble pile asteroids.
They set the terms of our ability to explore and inhabit it:
the physics of granular particles governs whether landers can stably progress, drills can penetrate, or loose regolith (extraterrestrial soil) can be transported and processed for construction or resource extraction \cite{Chaikin2020, Flahaut2023}.
Yet, granular processes still lack a predictive foundation: design is often empirical, 
relying on terrestrial intuition that breaks down in low gravity \cite{Wilkinson2005}. 
On a fundamental level,
the influence of (low) gravity on granular dynamics remains poorly understood \cite{Wilkinson2005,Sanchez2023,Shaebani2021,Cheng2024,Elekes2021}.
This gap becomes critical for devices that must work without fail
in hostile, poorly characterized conditions, 
less amendable to empirical testing.

The flow of material through an orifice is a basic of granular processing. 
When it interrupts spontaneously,
it is said to clog \cite{Zuriguel2020, Beverloo1961, Jenike1964, Walker1966}. 
Clogging happens if mesoscopic arch-like structures form within the packing, driving the entire system to a halt.
It can be mundane (salt stubbornly remaining in the saltshaker), 
as much as it can be dramatic (industrial failures).

\begin{figure*}[h!]
    \centering
\hspace{-5mm}
\includegraphics{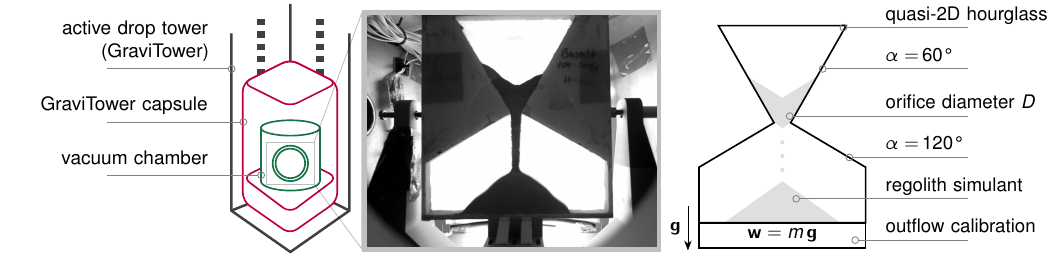}
    \caption{\label{fig:setup} 
\textbf{Experimental setup.}
A quasi-2D hourglass, placed in a vacuum chamber, is installed in the capsule of the active drop tower \acrfull{gtb} (\gls{zarm}, Bremen, Germany). A camera records the hourglass discharge through the chamber window; a typical image recorded is shown in the center panel,
featuring regolith simulant JSC-1A and an orifice diameter $D=$~\SI{10}{\mm}.
The quasi-2D hourglass is equipped with two hopper angles, $\alpha=$\,\SI{60}{\degree} (upper chamber here) and $\alpha=$\,\SI{120}{\degree} (lower chamber).
}
\end{figure*}

We focus on the effect of gravitational acceleration on the clogging probability during hopper outflow. 
Reducing gravity influences several factors relevant to clogging:
frictional forces (lowered \textit{via} lower shear along hopper walls)\cite{Lozano2012a},
reduced confinement \cite{Artoni2022} and packing density near the orifice \cite{DAngelo2022,Reiss2013,Yu2025}, 
and cohesion, which may dominate when particles' weight is reduced \cite{DAngelo2022,Sanchez2023,Yu2025}.
However, previous studies 
report contradictory findings regarding the effect of gravity on clogging probability,
indicating that gravity alone is not the appropriate control parameter for granular clogging.

Pioneer work by \citeauthor{Dorbolo2013} \cite{Dorbolo2013} report that gravity has no significant effect on clogging behavior.
This result is based on hypergravity experiments and mimicking reduced gravity using a 2D analogue and large metal spheres.
This was confirmed by \citeauthor{Arevalo2014} \cite{Arevalo2014,Arevalo2016}, who used 2D, cohesionless particle-based simulations across a wide gravity range ($10^{-3} g_E$ to $10 g_E$, where \red{$g_E = \SI{9.81}{\m\per\square\s }$}  is Earth gravitational acceleration):
they observed only a weak dependence of clogging on gravity, even in microgravity.
More experiments and simulations varying gravity,
although not primarily focused on clogging, also do not report any change in the clogging probability, $P_c$, with gravity
 -- both in low gravity ($g < g_E$) \cite{Brucks2008,Hofmeister2009} and hypergravity ($g > g_E$) \cite{LePennec1995,Mathews2016,Chung2008}.
These studies typically employ relatively low-cohesion granular materials (spherical glass or basalt, or natural sands with irregular but rounded shapes, larger than $\sim\SI{100}{\micro\meter}$).

On the contrary, parabolic flight experiments by \citeauthor{Reiss2014} \cite{Reiss2014} 
report an increased clogging probability under lunar gravity conditions, using regolith simulants. 
Experiments by \citeauthor{Ozaki2023} \cite{Ozaki2023}, conducted in a small centrifuge aboard the \gls{iss}, 
point in the same direction: 
although not focusing on clogging, 
more cohesive materials had to be excluded from the flow rate analysis
 -- notably lunar regolith simulant FJS-1 -- due to frequent clogging events \cite{Ozaki2023}.

Finally, a third line of evidence would suggest that lowering gravity
might actually \emph{reduce} clogging probability.
Two critical factors influencing $P_c$ are the particles packing density\cite{Endo2017,Koivisto2017} 
and confinement pressure \cite{Zuriguel2011,Lozano2012a,Unac2012}:
reducing pressure \cite{Zuriguel2011,Lozano2012b} or density \cite{Endo2017} near the outlet 
(for example by a well-placed obstacle \cite{Endo2017,Lozano2012a,Harada2022})
leads to reducing $P_c$.
In low gravity, experiments and simulations of hopper flow have shown that granular materials pack less densely near the outlet \cite{Reiss2013,Ozaki2023,Madden2025}, and the reduction in particle weight leads to lower confinement pressure. 
These factors together suggest that a decrease in gravity may result in a lower clogging probability.

Such contradictory results reveal a fundamental gap in our understanding of how reduced gravity affects granular flows \cite{Wilkinson2005, Chaikin2020, Sanchez2023, Dotson2024,Yu2025}.
This knowledge gap has contributed to operational challenges and, in some cases, 
failures during past space exploration missions \cite{Clegg2014,Jaffe1971,Lane2012,Apollo15MissionReport,Matson2010,Callas2015,Kerr2009,Kedar2017,Wippermann2020,Spohn2022}.
To avoid repeating mistakes of the past,
it is essential to understand how experimental results obtained under Earth gravity can be scaled to low gravity environments\cite{Jiang2018,Jiang2015,Chang2009,Madden2025,DAngelo2025c}.

We propose that the relevant control parameter is the granular Bond number, 
which quantifies the competition between cohesion and gravity.
We conduct hopper discharge experiments under Earth, Mars and Moon gravities, 
using the GraviTower active drop tower at the \gls{zarm} in Bremen, Germany \cite{Gierse2017}. 
The setup consists of a quasi-2D hourglass (see Fig.\,\ref{fig:setup}), 
placed in vacuum to ensure a homogeneous flow \cite{Wu1993}.
We select two hopper angles, $\SI{60}{\degree}$ and $\SI{120}{\degree}$,
positioned well below and above the known transition angle\cite{To2001, LopezRodriguez2019} $\alpha_c = \SI{75}{\degree}$. 
We use different lunar regolith simulants: crushed basalt, JSC-1A and LHS-2E (\textit{cf.}\ Methods). 
To characterize these materials, we propose an in-bulk definition of the granular Bond number, 
which enables to compare heterogeneous materials
and rescale results
across various gravitational accelerations.
Our results reveal a significant shift in clogging probability under reduced gravity.
As a scaling based only on gravity fails to capture the full complexity of granular flows,
we instead propose a scaling law using the granular Bond number, which incorporates both gravity and intrinsic material properties.
Using this approach, we reconcile previous results and
provide a unified framework to characterize granular flow in any gravitational environment.

\section*{Results}
\subsection*{The granular Bond number}

The main dimensionless number identifying gravity's effect on granular media is the granular Bond number, $\mathcal{B}$, 
defined as ratio of flow-inhibiting (cohesive) to gravity-driven forces,
\begin{equation}
\mathcal B = \frac{F_\text{cohesive}}{F_\text{gravity}}.
\end{equation}
It defines how \emph{flowable} or \emph{cohesive} a granular material is,
by quantifying the relevance of interparticle attractive forces.
Inspired by the study of fluids (where $\mathcal{B}$ is the ratio of flow-inducing hydrostatic pressure to flow-inhibiting surface tension and possibly yield stress \cite{Heitmeier2026}),
the granular version of $\mathcal{B}$ 
suffers from the difficulty to quantify cohesive interactions in granular media.
This is particularly true in highly heterogeneous materials.

\begin{figure}
\centering
\includegraphics{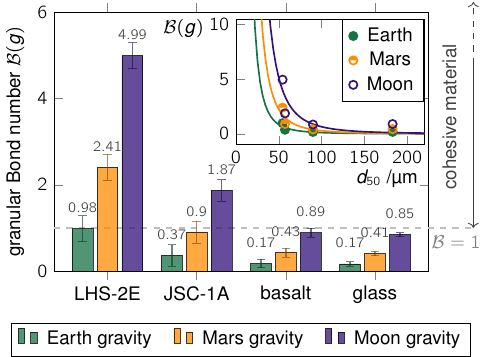}
\caption{\label{fig:bondnbrsvsg}
\textbf{Characterization of granular materials \textit{versus} gravity (Earth, Mars, Moon) by the granular Bond number.}
The values for $\mathcal{B}$ are based on Eq.\,\ref{eq:BondNbr}, using the $g_\text{L, M}$ recorded during experiments;
details on $\Upsigma_\text{t} $ are given in the Methods section;
error bars indicate the variability in $\Upsigma_\text{t} $.
The dashed line at $\mathcal{B}=1$ represents the theoretical limit at which material start to behave as cohesive.
In the inset, $\mathcal{B}$ is plotted \textit{versus} the median diameter ($d_{50}$) for the granular materials tested;
the theoretical estimates (lines) are fits for $\mathcal{B}\propto 1/d^3$.
}
\end{figure}

One possible approximation is to equate $\mathcal{B}$ with 
the ratio of theoretical \gls{vdw} forces between neighboring particles
to a particle's weight under Earth gravity \cite{Gupta2009,Hsu2018,Affleck2023}.
Then, if the \gls{vdw} forces are limited by the particles asperities, $\mathcal{B}\propto g/d^3$ (\textit{cf.}\ Fig.\,\ref{fig:bondnbrsvsg} inset) \cite{Gupta2009,Affleck2023}.
However, for highly heterogeneous granular materials with a broad particle size distribution
and highly non-spherical shapes,
akin to our regolith simulants,
such theoretical characterization might diverge significantly from reality.
Moreover, \enquote{cohesion} in granular media extends beyond \gls{vdw} forces alone \cite{Pouliquen2025}: 
friction, interlocking (geometrical cohesion \cite{Franklin2012,Aponte2024}), electrostatic interactions \cite{Grosjean2023Mechanism}, liquid bridges \cite{Badetti2018}
may all contribute to preventing particle separation.

We propose a new approach to rescaling Earth-measured data for predicting granular behavior in space, based on in-bulk characterization of interparticle interactions in granular materials, $F_\text{c}$.
The global resistance to flow due to interparticle interactions is
measured using the pressure drop during fluidization measurement \cite{Jaraiz1992,Valverde1998,Hsu2018,Soleimani2021,Affleck2023}.
A detailed description of the measurement procedure is provided in the Methods section; a brief overview is given below.

To fluidize a granular bed,
the drag force exerted by the fluidization air flow must balance the particles' weight per area, $mg/A$ ($m$ being the sample's mass, $A$ the bed's section area).
Once the pressure drop across the powder bed $\Delta P \geq mg/A$, the bed becomes fluidized. 
When measuring $\Delta p$ between fluidization and de-fluidization
($\Delta p_\text{up}$ and $\Delta p_\text{down}$, respectively), the hysteresis corresponds to the supplementary pressure needed for the air flow to break the cohesive bonds between particles.
This hysteresis has been called a granular tensile strength\cite{Valverde1998}, $\Sigma_\text{t}$, 
as it provides a quantification of the intrinsic global cohesion within the granular sample, $F_\text{c} \sim (\Delta p_\text{up} - \Delta p_\text{down}) =  \Sigma_\text{t} $.
Therefore, we define the granular Bond number as
\begin{equation}\label{eq:BondNbr}
\mathcal B = \frac{\Sigma_\text{t} }{\delta (mg/A)},
\end{equation}
where $\delta$ is the portion of the powder bed that is fluidized\cite{Affleck2023}.
Resulting $\mathcal{B}(g)$ are given in Figure\,\ref{fig:bondnbrsvsg} for our three lunar regolith simulants.
The same measurement is also conducted on spherical glass beads for the sake of comparison
(borosilicate glass particles of diameter 70-\SI{110}{\micro\meter} from \textit{Cospheric}).

We note that this measurement is intended as a means of quantifying resistance to flow in granular media, 
i.e., as a material characterization, not associated with any specific flow geometry or process,
\red{and conducted under Earth gravity.}
The normalization of $\Sigma_\text{t}$ is performed with respect to the setup in which it is measured,
so that the quantity $\Sigma_\text{t} / \delta(m/A)$ remains material-dependent rather than setup-dependent.

In Figure \ref{fig:bondnbrsvsg}, the effect of the shift from Earth to Mars and Moon gravity on the predominance of cohesive forces in granular materials is obvious:
when gravity is reduced, cohesive forces become predominant, and $\mathcal{B}$ strongly increases.
The limit above which materials start to behave like highly cohesive materials is around $\mathcal{B}=1$ (dashed line in Fig.\,\ref{fig:bondnbrsvsg}).
On-ground as on Mars and the Moon,
particle diameter has a strong impact on cohesion force, with $\mathcal{B}$ diverging at $d \to 0$ 
(Fig.\,\ref{fig:bondnbrsvsg} inset).
Note that $\mathcal{B} \gg 1$ happens at $d\approx \SI{20}{\micro\meter}$ in $g_{E}$,
while it is already reached at $d\approx \SI{40}{\micro\meter}$ in $g_{L}$:
the predominance of cohesive forces in low gravity results in a shift of cohesive behavior to larger diameters,
i.e., the same material appearing more cohesive \cite{Wilkinson2005,Sanchez2023,Yu2025}.

\red{
To summarize, our approach replaces gravity-only scaling by scaling with the ratio of cohesive forces to particle weight, enabling extrapolation from Earth-based measurements to different gravitational environments without additional reduced-gravity experiments.
}

\subsection*{Flowing and clogging}

\begin{figure*}[tbh!]
   \centering
\includegraphics{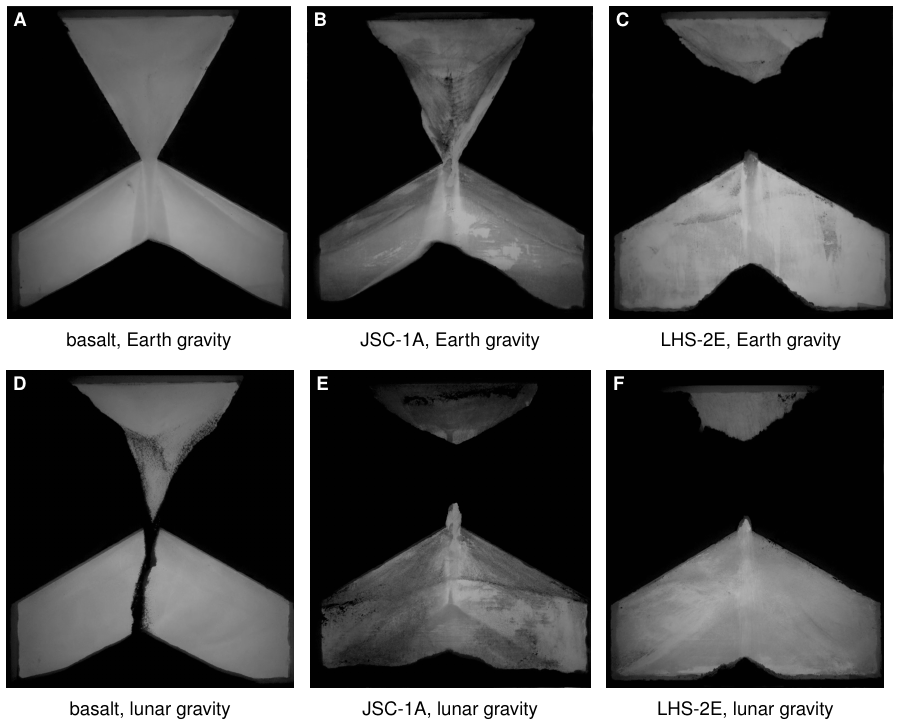}
   \caption{\label{fig:final_states}
\textbf{Last instant of hopper flow experiment for different lunar regolith simulants, under Earth and Moon gravitational accelerations.} 
Pictures are taken after \SI{2.3}{\s} in Earth gravity (A, B, C) and after \SI{2.3}{\s}, at the end of the lunar gravity phase (D, E, F), for the material indicated below each picture.
Note that in D, the basalt is still flowing at the end of the lunar gravity period.
The hopper geometry is the same in all frames: orifice width of \SI{10}{\mm}, hopper angle $\alpha=\SI{60}{\degree}$.
}
   \label{fig:final_states}
\end{figure*}

Clogging is a critical factor in granular hopper flow.
It results from local jamming of the granular material above the hopper orifice \cite{Zuriguel2005}.
In part due to its high relevance for industrial applications, 
extensive work has been done to predict clogging in hoppers;
important parameters include the orifice-to-particle size ratio, $D/d$,
and of the granular packing fraction, $\varphi$ \cite{Hafez2021,Zuriguel2014,Ozaki2023}.

To investigate the effect of gravity on clogging, we conduct experiments \red{under three different accelerations: 
Earth gravity $g_E$, close to Mars gravity, $g_M = 0.4 g_E$, and close to Moon gravity, $g_L = 0.19 g_E$
(note that these are measured values from our experiments),} 
on our three lunar regolith simulants 
(crushed basalt, JSC-1A, LHS-2E),
for hourglasses of orifice diameter $D$ from \SI{1}{\mm} to \SI{20}{\mm}. 
Let us use the case of $D= \SI{10}{\mm}$ to illustrate the different behaviors observed.

In Figure\,\ref{fig:final_states}, we see frames taken at the final instant of an experiment.
All hopper have a diameter $D=\SI{10}{\mm}$ and angle $\alpha=\SI{60}{\degree}$; the top row is for experiments conducted in Earth gravity, $g_E$, and the bottom row in Moon gravity, $g_L$.
The simulants display strikingly distinct behaviors: 
while basalt flows consistently in both Earth and lunar gravity, JSC-1A flows in $g_E$ but clogs in lunar gravity, and LHS-2E clogs even under Earth gravity conditions.

\begin{figure*}[tbh!]
\centering
\includegraphics{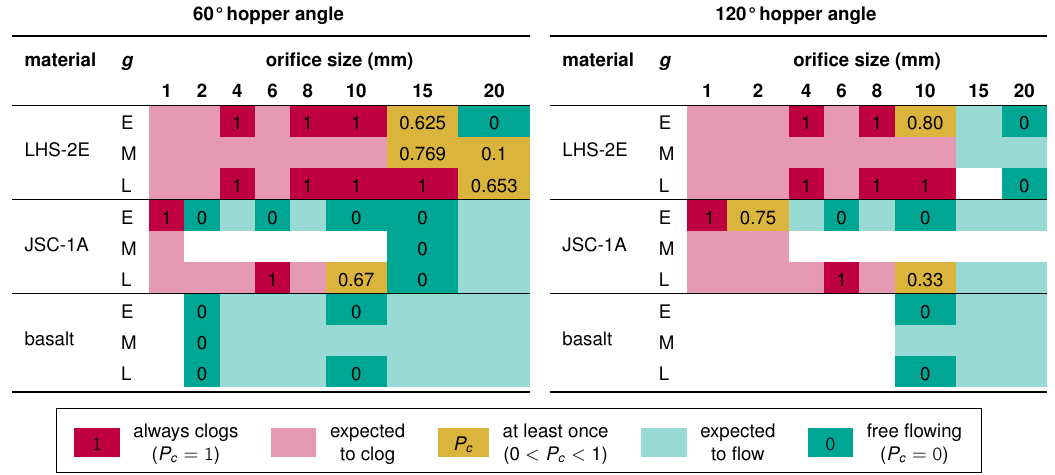}
\caption{\label{fig:CloggingProbas}
\textbf{Hopper clogging probability, $\bf P_c$:} free flowing ($P_c = 0$); clogged in at least one repetition ($0<P_c<1$); or clogged at each repetition ($P_c = 1$). Clogging statistics for all three materials studied (regolith simulants LHS-2E, JSC-1A and basalt beads), for varying orifice sizes $D\in[1,2,4,6,8,10,15,20]$\,\si{mm}, at funnel angles \SI{60}{\degree} (left) and \SI{120}{\degree} (right), and for Earth (E), Mars (M) and Moon (L) effective gravitational acceleration. Cells left white are untested; cells with lighter hue are untested but extrapolated from neighboring cases. }
\end{figure*}

Systematic measurements across different orifice sizes, shown in Figure\,\ref{fig:CloggingProbas},
confirm that reduced gravity increases clogging risk, 
with clogging already occurring at larger orifice diameter in lunar gravity than on Earth.
The critical orifice size at which clogging appears actually increases by approximately one order of magnitude between $g_E$ and $g_L$.
Between the extremes of free flow and clogging, a transition region is observed, where the behavior varies with each iteration.
Granular clogging being a statistical process, the clogging probability, $P_c$, decreases smoothly with increasing orifice width.

\begin{figure*}[h!]
\centering
\includegraphics{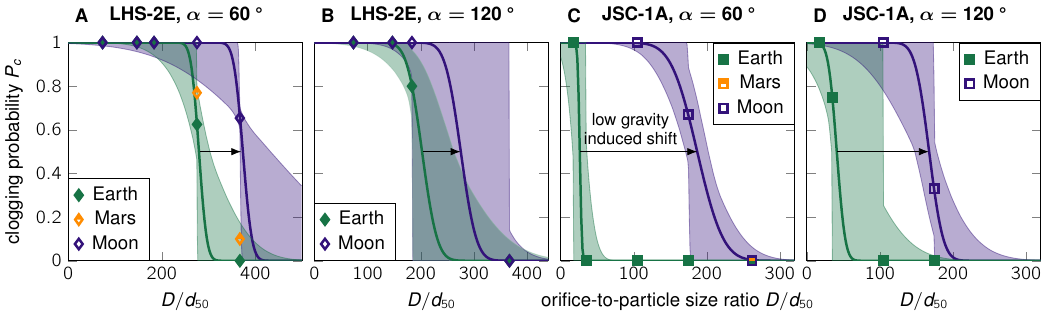}
\caption{\label{fig:CloggingCumulativeProba}
\textbf{Cumulative clogging probability as a function of orifice-to-particle size ratio. } 
The clogging probabilities are plotted for results in Earth and Moon effective gravitational acceleration,
for regolith simulants (A,B) LHS-2E and (C,D) JSC-1A, and hopper angles of (A,C) \SI{60}{\degree} and (B,D) \SI{120}{\degree}. 
Note the change in $x$-axis scale between (B) and (C). Black arrows represent the shift in clogging probability solely due to a change in effective  $g$.
The corresponding fit parameters and materials' particle size distributions 
are provided in the Methods section (Tab.\,\ref{tab:fitparamcloggingproba}).
Shaded areas indicate tolerance bands, which accept relative average deviations of $2/3$ per data point;
details in the Methods section.
}
\end{figure*}

For a single experiment, results can be plotted as the clogging probability against orifice-to-particle size ratio, $D/d$.
The median diameter ($d_{50}$) is used as representative of the particle diameter $d$.
The transition is modeled with a normal cumulative distribution function, 
expressed in terms of the error function as
\begin{equation}
y_{P_{50} , \sigma } (\nicefrac{D}{d_{50}}) = \frac{1}{2} \left( 1 - \text{erf} \left( \frac{\nicefrac{D}{d_{50}} - P_{50}}{\sigma \sqrt{2}} \right) \right).
\label{eq:erf}
\end{equation}

We plot the clogging probabilities for three different cases in Figure~\ref{fig:CloggingCumulativeProba}: 
JSC-1A for hopper angles $\alpha=\SI{60}{\degree}$ and $\alpha=\SI{120}{\degree}$, and LHS-2E, $\alpha=\SI{120}{\degree}$.
We present the $P_c(D/d_{50})$ graphs only for these selected cases with sufficient statistics.

For all four cases, lunar gravity entails a shift of the transition region (represented by a black arrow in Fig.\,\ref{fig:CloggingCumulativeProba}):
in $g_L$, clogging happens at larger $D/d_{50}$, i.e., clogging \emph{already} happens at larger orifice.

Comparing Figures \ref{fig:CloggingCumulativeProba}A and \ref{fig:CloggingCumulativeProba}B, or \ref{fig:CloggingCumulativeProba}C and \ref{fig:CloggingCumulativeProba}D, respectively,
we observe that the shift in clogging behavior varies little with hopper geometry.
However, the low-gravity-induced shift for LHS-2E (Fig.\,\ref{fig:CloggingCumulativeProba}C) -- a material with higher cohesion, see Fig.\,\ref{fig:bondnbrsvsg} -- is smaller than 
that of JSC-1A at equivalent hopper angle (Fig.\,\ref{fig:CloggingCumulativeProba}B).

This can be attributed to the intrinsic cohesion of LHS-2E already being significant under Earth gravity. 
In contrast, JSC-1A appears to reach the point where cohesion becomes dominant (typically around $\mathcal{B} \approx 1$)
within the tested gravity range, i.e., between $g_E$ and $g_L$.
This behavior suggests the presence of a material-dependent critical gravity threshold, beyond which granular properties, such as packing fraction and flow stability, undergo significant changes.

\subsection*{Clogging state diagram}

We have seen that reducing gravity increases $P_c$.
We now rationalize this observation
by including interparticle cohesion in our analysis.

In low gravity, cohesive interparticle forces, which might otherwise be insignificant compared to particles' weight, become predominant. 
This regime shift is captured by the granular 
Bond number, as defined above by measuring  in-bulk granular cohesion.
In other words, there exists a Bond number threshold that separates two regimes: 
one where interparticle cohesive forces dominate, 
and another one, where gravity dominates.

\begin{figure}
    \centering
    \includegraphics{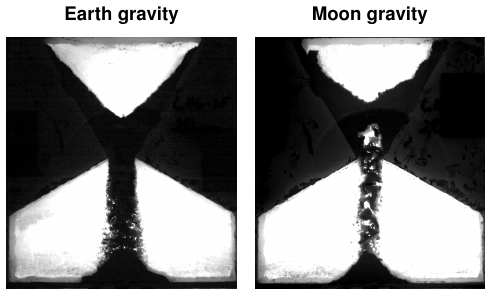}
    \caption{\label{fig:LHS_arc} \textbf{Illustration of qualitatively different flow in Earth \textit{vs.}\ Moon gravity. }
The same material (LHS-2E) and setup (hourglass with \SI{20}{\mm} orifice and hopper angle \SI{60}{\degree}) are used in Earth (left) and Moon (right). 
    }
\end{figure}

This behavior is qualitatively visible in Figure~\ref{fig:LHS_arc}: for the same material and geometry, clusters of particles form during free fall in low gravity, whereas the flow remains more homogeneous under Earth gravity.

\begin{figure}[h!]
    \centering
\includegraphics{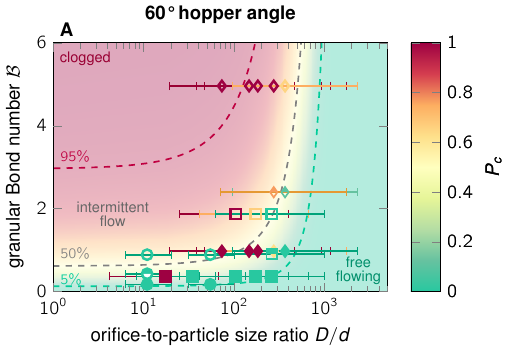}
\includegraphics{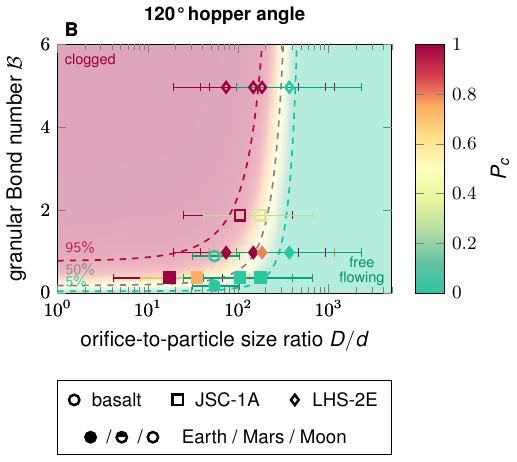}
    \caption{\label{fig:ClogStateRescaled}
\textbf{Clogging state diagram including range of polydispersity in granular systems.} 
Phase limits (colored background) are given by Eq.\,\ref{eq:state} (fit parameters for both datasets are given in Tab.\,\ref{tab:fitparastatediag}).
Each data point shows $D/d_{50}$ (including all materials and gravity conditions); horizontal error bars indicate the range of particle size distribution, from $D/d_{90}$ (lower bound) to $D/d_{10}$ (upper bound). 
}
\end{figure}

With this interpretation, Figure~\ref{fig:CloggingCumulativeProba} can be understood as showing the effect of the Bond number rather than gravity alone. 
Accordingly, we plot the clogging behavior as a function of orifice-to-particle size ratio and Bond number in Figure~\ref{fig:ClogStateRescaled}. 
To enable a more meaningful comparison between materials, the $d_{50}$ is used for scaling;
we discuss this rationale and its implications below.
A general trend emerges from this representation,
as all tested materials, in both Earth and Moon gravity, collapse onto a consistent behavior.

As shown in Figure~\ref{fig:CloggingCumulativeProba}, 
the clogging probability for a given material and gravitational acceleration (i.e., a given $\mathcal{B}$)
can be modeled by a cumulative distribution function (Eq.\,\ref{eq:erf}).
Lowering gravity, which effectively increases $\mathcal{B}$, shifts the probability to larger size ratios.
Thus, the parameter $P_{50}$, which describes the probability drop,
can be modeled as $\mathcal{B}$-dependent.

To model $P_{50}(\mathcal{B})$,
 we propose the function $P_{50}(\mathcal{B}) = a\cdot log_{10}(\mathcal{B}) + b$.
This form is chosen because it increases monotonically, reflecting the transition from clogged to free-flowing states (see e.g.\ Fig.\,\ref{fig:CloggingCumulativeProba}), and its slope decreases with $\mathcal{B}$,
as suggested by the data.
Moreover, the function remains unbounded, avoiding unphysical constraints at large $\mathcal{B}$,
where clogging may still occur due to cluster formation despite high size ratios.

Including $P_{50}(\mathcal{B})$ in Eq.\,\ref{eq:erf} leads to Equation~\ref{eq:state}, which describes the phase limits in our clogging state diagrams (Fig.\,\ref{fig:ClogStateRescaled}):
\begin{equation}
    \label{eq:state}
    P_{c}(\mathcal{B}o,\nicefrac{D}{d_{50}}) = \frac{1}{2} \left( 1- \text{erf} \left(\frac{\nicefrac{D}{d_{50}}-(a\cdot log_{10}(\mathcal{B}o) + b ))}{\sigma \cdot \sqrt{2}}\right)\right).
\end{equation}
The fit parameters ($a$, $b$, and $\sigma$) used are given in Tab.\,\ref{tab:fitparastatediag}.

\begin{table}[h!]
\sisetup{detect-all}
\small\normalfont\sffamily
\small\sansmath\sffamily
    \centering
    \caption{\label{tab:fitparastatediag}
\textbf{Fit parameters for clogging diagram phase limits.}
Numerical values of fit parameters for Eq.\,\ref{eq:state} ($a$, $b$ and $\sigma$), as plotted in Fig.\,\ref{fig:ClogStateRescaled}.
}
    \begin{tabular}{c c c c} 
	\toprule
     hopper angle 			& $a$ 			& $b$ 			& $\sigma$	\\
	\midrule
    \SI{60}{\degree} 		& 558.6 	& 117.1 	& 231.2		\\
    \SI{120}{\degree} 	& 205.9 	& 154.1 	& 79.4			\\
    \bottomrule
    \end{tabular}
\end{table}

A key challenge in predicting granular flow-behavior lies in selecting an appropriate characteristic length scale.
For monodisperse samples, the particle diameter provides an obvious choice; but for polydisperse materials, 
the choice becomes less evident. 
In particular, clogging behavior may be influenced disproportionately by either very large particles, which can block the orifice geometrically, or very small particles, where cohesion dominates, forming clusters that increase the effective particle size and increase the probability to form stable arches. 

While we use the median particle size, $d_{50}$, to normalize $D$, this may not be the most representative scale. 
To address this, we include horizontal error bars reflecting the particle diameter distribution from $d_{10}$ to $d_{90}$ in the state diagram in Fig.\,\ref{fig:ClogStateRescaled}. 
This approach reveals how polydispersity blurs the boundaries between flow regimes. 

Although working with real-life granular materials complicates modeling efforts compared to idealized systems (typically, monodisperse spherical glass beads), accounting for such complexity is essential to developing scaling laws relevant to practical applications. Our results show that, despite these complexities, the proposed Bond number framework remains robust across materials with diverse physical properties.
This highlights the need to consider not only geometric factors but also the evolving mechanical stability of the granular assembly under varying gravity conditions.

In summary, our results point to a clear physical mechanism underlying the observed behavior.
As previously demonstrated\cite{DAngelo2022,Yu2025}, the packing fraction associated with the jamming transition decreases when shifting from Earth gravity to reduced gravity.
Therefore, during granular hopper discharge, although reduced gravity leads to lower particle packing near the orifice -- as previously reported by \citeauthor{Reiss2013} and \citeauthor{Ozaki2023} \cite{Reiss2013,Ozaki2023} -- this does not necessarily correspond to a decreased clogging probability.
Instead, the granular packing is \emph{already} mechanically stable at reduced packing density, allowing clogging to occur even under looser configurations.

\section*{Discussion}

By testing multiple lunar regolith simulants under Earth, Mars and Moon gravitational accelerations, using an active drop tower, we demonstrate that reduced gravity significantly increases the clogging risk, challenging previous findings \cite{Dorbolo2013,Arevalo2014,Arevalo2016}. 
Across all materials and hopper geometries tested, we find a consistent shift toward higher clogging probability in low gravity, \red{despite the limited number of data points and repetitions.} 
Specifically, we observe a gravity-induced shift in the critical outlet size below which clogging occurs, by up to one order of magnitude.

Importantly, the extent of the gravity-induced shift depends on the intrinsic material properties.
We find that this shift is governed by cohesion, $\sigma_c$, measured \textit{via} pressure overshoot during fluidization 
(see Methods), 
More precisely, it is the relative balance between material cohesion and particles' weight under gravity that controls
its behavior.
This points directly to the Bond number, as defined earlier, as the relevant dimensionless parameter for scaling results between Earth and low gravity. 

We hypothesize that a material-dependent critical gravity threshold governs the transition in granular behavior, beyond which properties such as packing fraction and flow stability change significantly. Typically, LHS-2E exhibits substantial intrinsic cohesion already at Earth gravity, whereas JSC-1A appears to enter a cohesion-dominated regime within the tested gravity range, highlighting the interplay between gravity and material-specific cohesive forces.
The proposed framework is independent of planetary context and applies to any granular system in which body forces compete with cohesive interactions.

Previous studies reporting no significant effect of gravity on clogging probability \cite{Dorbolo2013,Arevalo2014,Arevalo2016} 
can be reconciled with experiments reporting contradictory findings \cite{Reiss2014,Reiss2013,Ozaki2023}, 
including the ones presented here, 
in light of our current understanding of the role of interparticle cohesive forces.
When studying the behavior of granular media, 
the straightforward interpolation of results from high to low gravity fails (where high and low correspond to above and below $g_E$, respectively).
It fails because at sufficiently low gravity, the balance of forces within the granular particles shifts:
cohesive interparticle forces become predominant over particles' weight, leading to a transition in the macroscopic flow behavior.
In low gravity, particles have been observed to adhere more strongly to each other, form clusters \cite{Love2014}, exhibit reduced packing efficiency \cite{DAngelo2022,Yu2025}, and show increased resistance to flow \cite{Yu2025}. 
As it is well established that cohesive particles have a higher tendency to clog during hopper discharge \cite{Umekage1998}, 
it follows that the same material, behaving more cohesively under low gravity, also exhibit an increased clogging probability.
This effect has been clearly observed in low-gravity experiments involving cohesive granular materials \cite{Reiss2014,Reiss2013,Ozaki2023}, particularly those containing fine particles where cohesion dominates over gravity. 
In contrast, experiments and simulations where interparticle cohesion is absent \cite{Arevalo2014,Arevalo2016} or minimal \cite{Dorbolo2013} understandably report no such increase in clogging.

\red{
Although reduced gravity also leads to lower packing density at the outlet \cite{Madden2025}, 
our experimental evidences clearly indicate that the resulting increase in the ratio of cohesive forces to particle weight is the dominant effect governing granular behavior across gravitational environments. 
This interpretation is further supported by the previous numerical studies \cite{Arevalo2014,Arevalo2016}
that found no significant change in clogging probability when only gravity was varied, 
despite the associated changes in packing density.
}

We note that for the \SI{120}{\degree} hopper,
the clogging probability is generally lower at same orifice size than for the  \SI{60}{\degree} hopper.
In the latter, the material is further accelerated by the steeper hopper angle (see Fig.\,\ref{fig:setup}),
which results in an increased clogging probability.
Interestingly, this goes in the direction of the \emph{slower-is-faster} effect \cite{Helbing2000,Pastor2015}
(which was \emph{not} observed in cohesionless, 2D simulations in low gravity \cite{Arevalo2014,Arevalo2016}):
instantaneously faster particles clog easier, resulting in a slower global outflow when time-averaged.
In our case, we can hypothesize that faster particles (or faster cohesion-induced clusters of particles)
in the \SI{60}{\degree} hopper 
result in higher pressure at the outlet, and therefore do experience a slower-is-faster effect.

Our results show that the transition to cohesion-dominated behavior depends not only on gravity but also on 
particle properties.
Other approaches have been proposed to define a granular Bond number, 
notably based on defining inertial numbers to compare the time-scales associated with intrinsic cohesion, gravity and imposed shear \cite{Azema2018}.
All current approaches conclude that
both particle properties (intrinsic) and extrinsic forces (gravity, shear) must be taken into account.
This has important implications for granular flow prediction in space applications, but also offers a model system for understanding highly cohesive materials, here on Earth. 
Such materials being used in numerous industries -- e.g., pharmaceuticals, food, manufacturing, construction,
advancing our understanding of highly cohesive granular materials promises large ecological and economic benefits.

\section*{Methods}

\subsection*{Partial gravity platform}
\begin{figure}[h!]
    \centering
\includegraphics{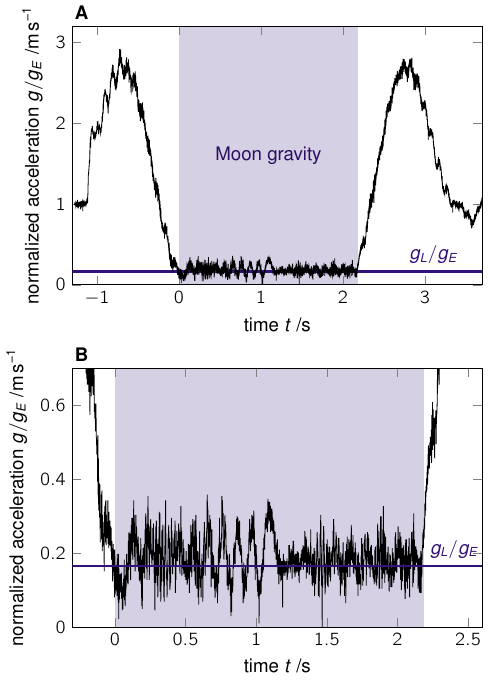}
    \caption{\label{fig:g-Moon}
\textbf{Effective acceleration during a typical GraviTower Moon gravity test.}
(A)~Effective acceleration, $g$, during a full GraviTower cycle, normalized by Earth gravity, $g_E$. The phase in lunar gravity, $g_L$, (blue shaded background) is framed by an acceleration and a deceleration phase.
(B)~Closeup of $g$ during the lunar gravity phase (blue shaded background); true lunar gravity (normalized by Earth gravity, $g_L/g_E=0.165$) is given for reference.
The $y$-axis scaling in (B) is the same as Fig.\,\ref{fig:g-Mars}B to ease comparison.
    }
 \end{figure}

\begin{figure}[h!]
    \centering
\includegraphics{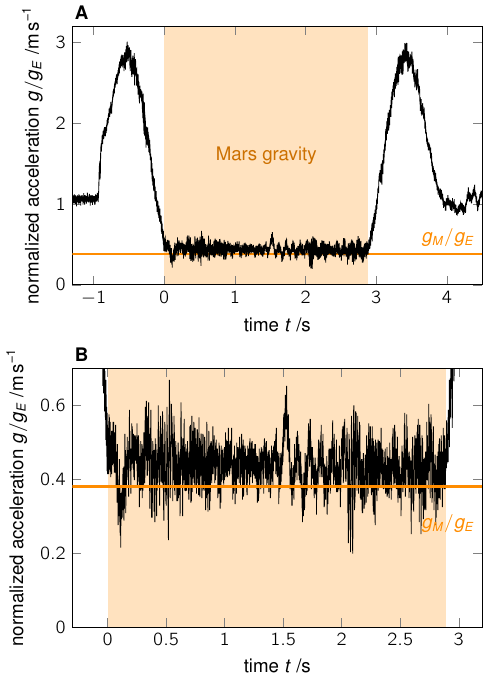}
    \caption{\label{fig:g-Mars}
\textbf{Effective acceleration during a typical GraviTower Mars gravity test.}
(A)~Effective acceleration, $g$, during a full GraviTower cycle, normalized by Earth gravity, $g_E$. The phase in martian gravity, $g_M$, (orange shaded background) is framed by an acceleration and a deceleration phase.
(B)~Closeup of $g$ during the martian gravity phase (orange shaded background); true Mars gravity (normalized by Earth gravity, $g_M/g_E=0.38$) is given for reference. 
The $y$-axis scaling in (B) is the same as Fig.\,\ref{fig:g-Moon}B.
    }
 \end{figure}

Lunar and martian gravity are generated with the \acrfull{gtb} at the \acrfull{zarm} (Bremen, Germany).
The GraviTower is an active drop tower, meaning that the capsule is not dropped but continuously accelerated \cite{Konemann2015,Gierse2017,Gierse2022}.
This controlled acceleration makes it possible to create partial gravity in the experimental capsule.
The acceleration profiles over one flight are shown in Figures~\ref{fig:g-Moon} and \ref{fig:g-Mars}, for Moon and Mars, respectively.
The full acceleration profile for a single flight is shown in Figs.~\ref{fig:g-Moon}A and\ref{fig:g-Mars}A. 
The flight begins with an upward acceleration phase, during which the experimental capsule (Fig.~\ref{fig:setup}A) reaches a peak acceleration of approximately $3g_E$. This is followed by a stable phase of reduced gravity, during which the capsule experiences an acceleration equivalent to lunar (martian) gravity for approximately \SI{2.3}{\second} (\SI{2.9}{\second}).
The flight concludes with a deceleration phase, which mirrors the launch acceleration.

In this experimental campaign using the GraviTower, the average effective gravitational acceleration, $g$,
achieved during individual flights varied between $0.186 g_E$ and $0.207 g_E$ for the lunar flights,
and between $0.405 g_E$ and $0.408 g_E$ for the martian flights.
The typical standard deviation for a single flight in both gravity settings is $0.07 g_E$.
As shown in the closeups in Fig.\,\ref{fig:g-Moon}B and Fig.\,\ref{fig:g-Mars}B,
a dominant oscillation is observed 
at a frequency of $9.5 \pm \SI{0.5}{\Hz}$.
This oscillation is the main contributor to the standard deviation;
\red{it is close to the resonance frequency of the pulley system of the active drop tower, 
which is likely the reason for this oscillation.}

While vibrations can enhance the flow of granular materials \cite{Janda2009}, 
the increased clogging probability and significant outflow rate reduction remain clearly observed in our data. 
Therefore, the oscillatory component of $g$ is not considered critical for the interpretation of our results;
it may suggest that the issues discussed here are even more pronounced under ideal, vibration-free lunar gravity conditions.

\subsection*{Experimental setup}
The hourglasses used in this experiment are machined from aluminum, with glass windows are glued to the front and back. 
Internal pressure is regulated inside the hourglasses \textit{via} a borehole covered by a wire mesh (mesh 120, $\SI{125}{\micro\meter}$ spacing) to prevent particle loss. 
Each hourglass measures \SI{20}{\centi\meter} in length, \SI{18}{\centi\meter} in width, and \SI{1}{\centi\meter} in thickness.
Many versions of the hourglass with orifice sizes, $D$, in the range from \SI{1}{\mm} to \SI{20}{\mm} have been produced.
The two compartments of the hourglass have hopper angles of \SI{60}{\degree} and \SI{120}{\degree} (see Fig.\,\ref{fig:setup}),
so both can be tested with the same hourglass.

\begin{figure}[h!]
\centering
\includegraphics{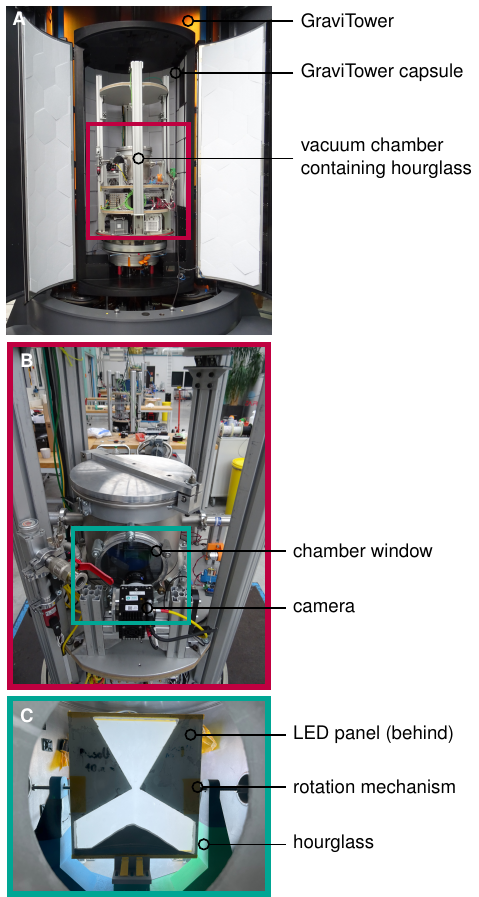}
\caption{\label{fig:setup-inmethods}
\textbf{Experimental setup integrated in the GraviTower.} 
(A)~Capsule with the experimental setup, placed in the GraviTower.
(B)~Vacuum chamber with the camera and external support systems.
(C)~View of the hourglass taken from the vacuum chamber window (position of the camera). 
}
\end{figure}

The hourglass is placed inside a vacuum chamber with a window.
A camera records the hourglass through the vacuum chamber window at \SI{1000}{\fps}, as shown in Figure~\ref{fig:setup-inmethods}. 
The experiment is conducted at least three times with every set of parameters (material, orifice size, hopper angle, gravity).

Due to spacial limitations, the vacuum pump is not connected during GraviTower flights.
The experiment is subsequently conducted at a typical pressure of $5.4 \pm \SI{0.1}{\milli\bar}$.
The pressure is nearly constant throughout each experiment, with a leakage rate of \SI{3e-4}{\milli\bar\per\second}.

\subsection*{Materials}

Two regolith simulants have been tested and compared to a sample of crushed basalt (particle size distribution ranging from \SIrange{10}{300}{\micro\meter}). 
The simulants are LHS-2E~\cite{LHS2E} (commercially available simulant proposed by space resource technologies) and JSC-1A~\cite{JSC1A} (provided by the Johnson Space Center, Texas, USA).

\begin{figure*}[bth!]
\centering
\includegraphics{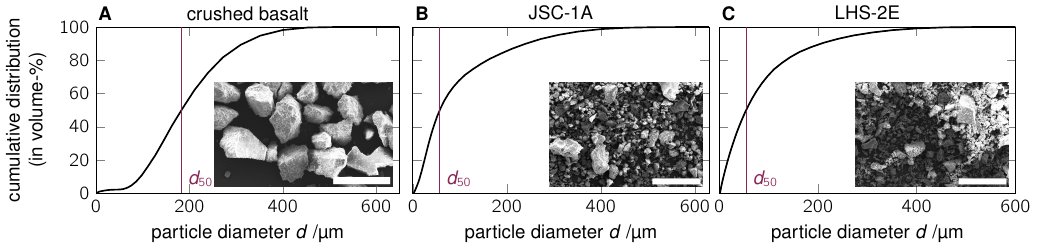}
\caption{\label{fig:cumulativeparticlesizedistrib}
\textbf{Cumulative particle size distributions of regolith simulants.}
(a)~Crushed basalt ($d_{50}=$\,\SI{183}{\micro\m});
(b)~JSC-1A ($d_{50}=$\,\SI{57.3}{\micro\m});
(c)~LHS-2E ($d_{50}=$\,\SI{54.5}{\micro\m}).
Microscopies of the materials are given in inset; scale bars all represent \SI{500}{\micro\meter}.
Data and microscopies reproduced with permission from Ref.\ [\citenum{Lobach2024}].
}
\end{figure*}

The batches of materials used in our analysis were specifically characterized. 
Cumulative particle size distribution for the three materials, as well as the median diameters, 
are given in Fig.\,\ref{fig:cumulativeparticlesizedistrib}.

\subsection*{Granular Bond number}

The granular Bond number, $\mathcal{B}$, as defined in Eq.\,\ref{eq:BondNbr},
quantifies the ratio of interparticle (intrinsic) cohesive forces \textit{versus} particles' weight (extrinsic).
It provides a threshold to distinguish the dominant forces in the granular media:
if $\mathcal{B} \gg 1 $, interparticle cohesion dominates;
if $\mathcal{B} \ll 1 $, particles' weight dominates.

Measuring or calculating cohesive forces in granular media can prove difficult, 
as theoretical predictions require simplifying assumptions regarding particle morphology, size distribution, surface properties, among others. 
While the \enquote{pull-off force} between individual particles can be measured, for example using \gls{afm} \cite{Colbert2015}, 
such methods only probe specific particle pairs and may not be representative of the entire granular sample. 
This limitation is particularly relevant for natural granular materials, which are typically heterogeneous in both size and composition. 
For such materials, an in-bulk measurement approach is more appropriate, as it captures the collective effect of interparticle forces and provides an effective value of the cohesive force representative of the entire packing.

One approach to quantifying the effective in-bulk cohesion within a granular sample is through the pressure drop measurement
across a fluidized bed. By gradually increasing (or decreasing) the airflow through the granular sample, 
while recording the pressure drop across the bed, \red{$\Delta p$,}
one can determine the onset of fluidization (or defluidization). 
At the onset of fluidization, the upward drag force exerted by the air balances the weight of the granular bed. 
A characteristic pressure overshoot is typically observed just before the material becomes fully fluidized or suspended. 
This overshoot reflects the additional force required to overcome interparticle cohesion and initiate the transition from a static to a fluidized state. 
The magnitude of this overshoot -- recorded as the difference between the increasing pressure (\red{$\Delta p_\text{up}$}) and the decreasing pressure (\red{$\Delta p_\text{down}$}) -- has been referred to as the granular tensile strength \cite{Valverde1998} 
and can serve as an effective proxy for the in-bulk cohesive force, $F_\text{cohesive}$, in various granular systems.

Recently, the overshoot in pressure drop measurements have been employed to approximate an experimental granular Bond number, providing an in-bulk and repeatable method for characterizing cohesion in granular media \cite{Hsu2018, Soleimani2021, Affleck2023}.
While this approach offers a practical alternative to particle-scale force measurements, it remains based on several assumptions.
In particular, most formulations rely on idealized particle morphologies (typically assuming spherical particles) to estimate the coordination number and other microstructural parameters.

Our samples are highly polydisperse and composed of non-spherical grains (see Fig.\,\ref{fig:cumulativeparticlesizedistrib}), 
making traditional assumptions inappropriate. 
To avoid introducing such assumptions, we propose a definition of the granular Bond number, $\mathcal{B}$, based solely on experimentally accessible quantities extracted from pressure drop measurements. This formulation,
particularly inspired by the approach of \citeauthor{Affleck2023} \cite{Affleck2023},
enables direct comparison between materials, irrespective of morphology, size distribution, and so on.

We define the effective cohesive stress in the sample as the pressure hysteresis observed during fluidization:
$F_\text{cohesive} \sim (\Delta p_\text{up} - \Delta p_\text{down}) =  \Sigma_\text{t} $.
This cohesive stress is normalized by the effective weight per unit area of the fluidized portion of the sample, yielding
Eq.\,\ref{eq:BondNbr}:
$\mathcal{B} = \Sigma_\text{t} / \delta (mg/A) $.
Here, $m$ is the total sample mass, $g$ the gravitational acceleration, $A$ the cross-sectional area of the bed, and $\delta$ the fraction of the sample mass that becomes fluidized. The latter is estimated from the steady-state pressure drop during fluidization, \red{$\Delta p_\text{SS}$, such that $\delta = \Delta p_\text{SS} / (mg/A)$}.

This formulation of $\mathcal{B}$ provides a bulk, assumption-free metric for cohesion that is directly grounded in experimentally observed behavior.
\red{Importantly, $\mathcal{B}$ can be evaluated for arbitrary gravitational environments from a single Earth-based measurement, 
valid for heterogeneous materials, without requiring additional reduced-gravity testing.}

\begin{figure}[h!]
\centering
\includegraphics{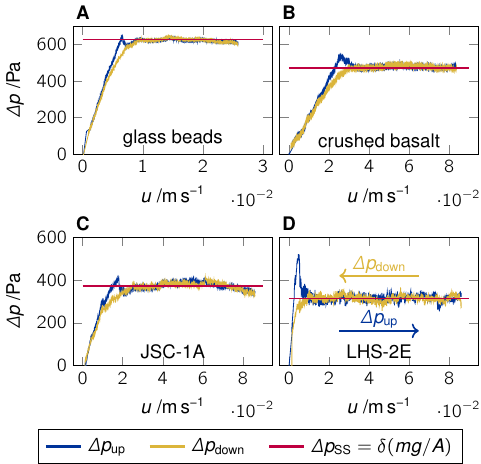}
\caption{\label{fig:pressuredrop}
\textbf{Pressure drop measurements.}
The pressure drop across the granular bed,
$\varDelta p$, is measured using an Anton Paar MCR-102, 
\red{when going down ($\varDelta p_\text{down}$) then up ($\varDelta p_\text{up}$) in air flow velocity, $u$.}
A slow paddle rotation (\SI{5}{\per\minute}) is added to ensure homogeneous fluidization across all materials. 
}
\end{figure}

Pressure drop measurements are conducted on an Anton Paar Modular Compact Rheometer MCR-102, 
fitted with a camera that records the volume occupied by the sample 
to access its global packing density, $\varphi$ (full description of setup is available elsewhere \cite{DAngelo2025}).
Typical pressure hysteresis loops are exhibited in Fig.\,\ref{fig:pressuredrop}.
As all materials tested are cohesive and polydisperse, we add a slow rotation of a 
four-blades stirrer (or paddle) at rotation rate of $n=\SI{5}{\per\minute}$
to aid homogeneous fluidization and make the measurements more reproducible.
\red{Note that this will slightly decreases the measured cohesion; therefore, it is applied consistently to all samples to allow meaningful comparisons. 
For example, for the reference glass beads, measurements performed with and without agitation 
($n=\SI{5}{\per\minute}$ and $n=\SI{0}{\per\minute}$, respectively) 
yield $\Sigma_t = \SI{103}{\pascal}$ and $\Sigma_t = \SI{126}{\pascal}$, respectively.
A similar comparison cannot be performed for the other materials, as their cohesion is sufficiently high that homogeneous fluidization cannot be achieved without agitation, due to channel formation and partial fluidization of the bed.
}

From the measured hysteresis loops, we calculate $\mathcal{B}$ for each sample as a function of gravity 
(Fig.\,\ref{fig:bondnbrsvsg}). For reference, we also include the values of $\mathcal{B}$ obtained for a reference sample of borosilicate glass beads (spherical particles with diameters in the range \SIrange{70}{110}{\micro\meter} from \textit{Cospheric}).
The corresponding Bond numbers at Earth gravity ($g = g_E$) are summarized in Table~\ref{tab:granularmaterials},
as well as the low gravity $\mathcal{B}$, calculated for our experimental $g$. The average particle diameter $d$ used for normalization is the median diameter, $d_{50}$, also reported in Table~\ref{tab:granularmaterials}.

It is important to note that the zero-air-flow point ($\Delta p=0$)
is critical in this analysis, as it anchors the effective weight-based normalization ($\delta$) and therefore defines the baseline for evaluating the cohesive contribution.

\begin{table}[h!]
\sisetup{detect-all}
\small\normalfont\sffamily
\small\sansmath\sffamily
    \centering
    \caption{\label{tab:granularmaterials}
\textbf{Material parameters for granular materials studied.}
For each material, we give the median diameter, $d_{50}$, the cohesion $\Upsigma_\text{t}$ as measured by pressure drop measurement, and granular Bond numbers for Earth, Mars and Moon gravity, $\mathcal{B}(g_\text{E})$, $\mathcal{B}(g_\text{M})$ and $\mathcal{B}(g_\text{L})$, respectively.
}
    \begin{tabular}{c S[table-format=3.2] S[table-format=3.2] S[table-format=3.2] S[table-format=3.2] S[table-format=3.2]} 
	\toprule
    material &{$ d_{50}$ /\si{\micro\m}}		&{$\Upsigma_\text{t}$ /\si{\pascal}}	& {~~$\mathcal{B}(g_\text{E})$}& {~~$\mathcal{B}(g^\text{exp}_\text{M})$	}& {~~$\mathcal{B}(g^\text{exp}_\text{L})$}  \\ 
	\midrule
    glass 					&	90	&	103.3		&	0.167	&	0.410	&	0.848	\\
    basalt					&	183	&	83.3 		&	0.175	&	0.430	&	0.890	\\
    JSC-1A 				&	57.3	&	136.7		&	0.368	&	0.905	&	1.872	\\
    LHS-2E 				&	54.5	&	306.7		&	0.974	&	2.397	&	4.960	\\
    \bottomrule
    \end{tabular}
\end{table}

\subsection*{Clogging probability analysis}

Every recorded video is investigated for clogging. 
An iteration is counted as clogged, if the material stops falling into the bottom compartment and if the orifice is blocked by material in the upper chamber. 
The number of clogged iterations were counted and divided by the full number of iterations.
The resulting number is interpreted as clogging probability, $P_c$.
The probabilities were calculated from a minimum of three iterations.

The clogging probability is plotted against the orifice-to-particle size ratio in Fig.\,\ref{fig:CloggingCumulativeProba}.
For better visualization of the difference between lunar gravity und earth gravity, a normal error function is fitted by a least-square-fit to the data points. 
The tolerance bands accept a maximum summed deviation of data and fit of $N \cdot \nicefrac{2}{3}$, where $N$ is the number of data points. 
\red{The number of data points and repetitions limits the precision of this fit, thus it should primarily be regarded as a visual support for the reader.}
The fit parameters are shown in Table~\ref{tab:fitparamcloggingproba}.

In the fit procedure for Figs.\,\ref{fig:CloggingCumulativeProba} and \ref{fig:ClogStateRescaled}, 
we do not include
the number of measurements per parameter combination.
This choice is made
because the variability between iterations is very large in the transition region ($0<P_c<1$), 
while it is small in the constant regions ($P_c=1$ and $P_c=0$).
Thus, albeit a larger number of repetitions, 
assuming a higher precision in the transition region
would lead to underestimating the width of the transition region.

\begin{table}[h!]
\sisetup{detect-all}
\small\normalfont\sffamily
\small\sansmath\sffamily
    \centering
    \caption{\textbf{Fit parameters for clogging probability analysis.}  
used in Fig.\,\ref{fig:CloggingCumulativeProba} based on Eq.\,\ref{eq:erf}
hopper corresponds to the hopper angle use din the experiments analyzed.
    \label{tab:fitparamcloggingproba}}
    \begin{tabular}{c c c c c c c } 
	\toprule
 									&& \multicolumn{2}{c}{Earth} &&  \multicolumn{2}{c}{Moon}  \\ 
    material (hopper)		&& $P_{50,\text{E}}$	& $\sigma_\text{E}$ && $P_{50,\text{L}}$ & $\sigma_\text{L}$ \\ 
	\midrule
    JSC-1A (\SI{60}{\degree}) 		&& 26.2 		& 2.3 		&&	186.9 		& 28.1		\\
    JSC-1A (\SI{120}{\degree})		&& 41 			& 9.1 	    &&	167.4		& 15.2		\\
    LHS-2E (\SI{60}{\degree})   	&& 279.5 		& 13.3  	&&	373.4		& 15.3		\\
    LHS-2E (\SI{120}{\degree})   	&& 202.5 		& 22.5  	&&	275.4		& 23.1		\\
    \bottomrule
    \end{tabular}
\end{table}

\def\bibfont{\footnotesize}
\printbibliography

\section*{Acknowledgements}
We thank Jan Lob\"ach for analyzing the simulants as shown in Figure~\ref{fig:cumulativeparticlesizedistrib}.
O.~D'A.~warmly thanks Naomi Murdoch and Mazi Jalaal for critical reviewing of the work.

This work was supported by the German Aerospace Center (DLR) Space Administration with funds provided by the German Federal Ministry for Economic Affairs and Climate Action (BMWK) under grant numbers 50WM2342A and 50WM2342B.
O.~D'A.~acknowledges financial support from the French National Centre for Space Studies (CNES) under the CNES fellowship 24-357.

\section*{Data Availability}
The original data, in the form of the videos recorded,
are publicly available 
on the Zenodo repository 15577225, see Ref.\ [\citenum{Gaida2025dataset}].

\section*{Author contributions statement}
\textbf{O.G.}: Investigation (lead); Data curation; Formal analysis; Visualization; Writing -- original draft
\textbf{O.D'A.:} Conceptualization (supporting); Funding acquisition; Investigation; Formal analysis; Visualization; Writing -- original draft.
\textbf{J.K.}: Conceptualization (lead); Funding acquisition; Resources; Supervision; Project administration; Writing -- review \& editing.

\section*{Competing interests statement}
The authors declare no competing interest.

\end{document}